%% file: egas_cpimcxxx.tex
\documentclass[cpp,fleqn]{w-art}
\usepackage{times}
\usepackage{w-thm}
\usepackage[utf8]{inputenc}
\usepackage{amsmath}
\usepackage{amssymb}
\usepackage{graphicx}
\usepackage{tikz,pgfplots}
\usetikzlibrary{arrows,decorations.pathreplacing,backgrounds,positioning,fit,calc,through,shapes} 

\usepackage[english]{babel}

\usepackage{mathtools}
\usepackage{braket}
\usepackage{booktabs}
\input{mydefinitions.tex}

\begin{document}
\DOIsuffix{theDOIsuffix}
\Volume{42}
\Issue{1}
\Month{01}
\Year{2013}
\pagespan{1}{}
\Receiveddate{}
\Reviseddate{}
\Accepteddate{}
\Dateposted{}



\title[CPIMC results for the electron gas at strong degeneracy]{Towards ab initio thermodynamics of the electron gas at strong degeneracy}


\author[T. Schoof]{T. Schoof}
\author[S. Groth]{S. Groth}
\author[M. Bonitz]{M. Bonitz\footnote{Corresponding
     author: e-mail: {\sf bonitz@physik.uni-kiel.de}, Phone: +49\,431\,8804122,
     Fax: +49\,431\,8804094}}
\address[]{Christian-Albrechts-Universität zu Kiel, Institut für Theoretische Physik und Astrophysik, Leibnizstraße 15, 24098 Kiel, Germany}
\newcommand{\todo}[1]{\textcolor{red}{\underline{#1}}}
\newcommand{\eq}[1]{Eq.~(\ref{#1})}
\newcommand{\nn}{\nonumber}
\newcommand{\e}[1]{\mathrm{e}^{#1}}
\newcommand{\eqsand}[2]{Eqs.~(\ref{#1}) and~(\ref{#2})}
\newcommand{\cc}{{\cal C}}
\newcommand{\mean}[1]{\langle #1 \rangle}
\newcommand{\tc}{T_{\cal C}}
\newcommand{\dc}{\delta_{\cal C}}
\newcommand{\ii}{\mathrm{i}}
\newcommand{\thc}{\theta_{\cal C}}
\newcommand{\intc}[1]{\int_{\cal C}\mathrm{d}{#1}\;}
\newcommand{\intlim}[3]{\int_{#1}^{#2}\mathrm{d}{#3}\;}
\newcommand{\mret}{{\mathrm{ret}}}
\newcommand{\madv}{{\mathrm{adv}}}

\begin{abstract}
Recently a number of theoretical studies of the uniform electron gas (UEG) at finite temperature have appeared that 
are of relevance for dense plasmas, warm dense matter and laser excited solids and thermodynamic density functional theory simulations. In particular, restricted path integral Monte Carlo (RPIMC) results became available which, however, due to the Fermion sign problem, are confined to moderate quantum degeneracy, i.e. low to moderate densities. We have recently developed an alternative approach---configuration PIMC [T. Schoof {\em et al.}, Contrib. Plasma Phys. {\bf 51}, 687 (2011)] that allows one to study the so far not accessible high degeneracy regime. Here we present the first step towards UEG simulations using CPIMC by studying implementation and performance of the method for the model case of $N=4$ particles. We also provide benchmark data for the total energy.
\end{abstract}
\maketitle                   

\section{Introduction}\label{s:intro}
Thermodynamic properties of quantum degenerate electrons are vital for the description of matter at high densities, such as plasmas in compact stars or planet cores, as well as in laser fusion experiments at the National Ignition Facility (NIF), e.g.~\cite{lindl_04, hurricane_nif14}  or for the imploding z-pinch Liners at Sandia National Lab~\cite{hanson_13}. Besides, the electron component is of crucial importance for understanding the properties of atoms, molecules and real materials. Since exact wave function based methods for solving the many-electron problem are hampered by an exponential slowing down with increasing number of electrons, e.g. \cite{hochstuhl_jcp11}, many-body methods are of central importance, e.g. \cite{mahan-book, bonitz_cpp13}. However, these methods have a limited accuracy determined by the used approximation and are usually limited to weak or moderate coupling.
Alternatives, therefore, have been first principle simulations such as path integral Monte Carlo (PIMC), e.g. \cite{ceperley95rmp}, however, in the case of fermions they suffer from the fermion sign problem (FSP). It prevents direct 
fermionic simulations, e.g. \cite{filinov-etal.00pla,FiBoEbFo01} at strong degeneracy, $\chi = n \lambda^3_{DB} \gg 1$, 
where $\lambda_{DB}^2=h^2[2\pi m k_BT]^{-1}$ denotes the thermal DeBroglie wave length and $n$ is the density.
The FSP can be ``avoided'' by performing ``restricted'' PIMC (RPIMC) simulations using fixed nodes, 
 e.g. \cite{mil-pol} and references therein, but their error is difficult to assess. 
Recently finite temperature RPIMC (DPIMC) simulations have also been performed for the uniform electron gas \cite{brown_prl13} (\cite{filinov_14}), but due to the FSP, reliable results are, most likely, restricted to moderate densities, $r_s \gtrsim 1.5$ [$r_s={\bar r}/a_B$, where ${\bar r}$ is the mean interparticle distance, $n^{-1}= 4\pi {\bar r}^3/3$ and $a_B$ the Bohr radius] and temperatures above $\Theta=k_BT/E_F=0.0625$, where $E_F$ is the Fermi energy. 
However, this leaves out the high-density range that is of high importance, e.g. for deuterium-tritium implosions at NIF where mass densities of $400 $ gcm$^{-3}$ have recently been reported \cite{hurricane_nif14}, corresponding to $r_s\approx 0.24$.
%
%
To bridge the gap between the known analytical result for the ideal Fermi gas and the RPIMC data, 
recently  several fits have been proposed \cite{brown_prb13, karasiev_prl14} but they also require reliable 
first-principle data at low $r_s$. 
We have recently demonstrated \cite{schoof_cpp_11} that a suitable approach to PIMC simulations at high degeneracy is given by simulations in Slater determinant space (configuration PIMC, CPIMC). For the model of fermions in a harmonic oscillator we could report CPIMC results that are uncaccessible for DPIMC and are essentially complementary with respect to the FSP \cite{simon_springer14}. We are presently adapting this approach to the uniform electron gas and here present first results. For illustration we analyze a small system of $N=4$ spin polarized fermions as this allows for comprehensive tests of the behavior of the sign as a function of density, temperature and basis size and to compare to exact diagonalization results.

\section{Configuration path integral Monte Carlo (CPIMC)}\label{s:cpimc}
The thermodynamic properties of a quantum mechanical many-body system in equilibrium are fully determined by the density operator $\rho$ which, in the canonical ensemble, is given by 
${\hat \rho} = Z^{-1}e^{-\beta \Ham}$,
with the inverse temperature $\beta$, the Hamiltonian $\Ham$ and the partition function $Z=\Tr {\hat \rho}$. As the internal energy and many other thermodynamic quantities can be derived from $Z$ we are looking for a numerically tractable expression. 
The usual approach is to expend the trace in the coordinate representation, decomposing ${\hat \rho}$ into a product of $M$ density operators, each defined at an $M$-times higher temperature, and approximating these using the Trotter formula or a higher order scheme. This leads to the well-known path integral formulation of the partition function. Because the many-body coordinate states are simple product states, they do not fulfill the appropriate particle statistics for fermions or bosons, and one has to apply the (anti-)symmetrization operator to at least one of the states. For fermions this introduces a sign change for odd permutations of particles making the calculation of the integral exponentially difficult with increasing particle number and inverse temperature---this is the fermion sign problem.
\par
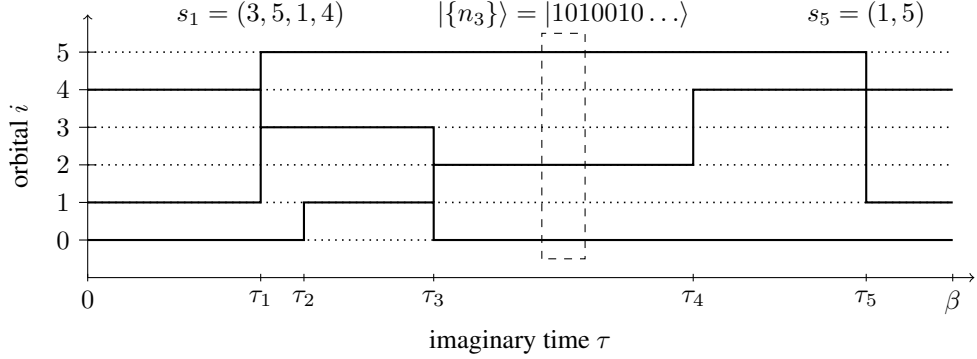
\begin{figure}[t]
\centering{\input{original_to_wa_picture_path.tikz.tex}
\caption[Path in kink picture]{Possible path $\ket{\occconfig{n}}(\tau)$ in imaginary time of three particles in six orbitals in the kink picture. Each kink $s$ represents either a one- or a two-particle excitation.\label{fig:sample_path}}}
\end{figure}

	
The basic idea of CPIMC is to use, for evaluation of the trace, an arbitrary complete orthonormal set of basis functions that fulfills the correct symmetry under particle exchange. We will use occupation number (Fock) states 
\begin{equation}
\ket{\occconfig{n}}:=\ket{n_1n_2\ldots},\quad n_i=0,1.
\end{equation}
In Ref. \cite{schoof_cpp_11} we derived the expression for $Z$ in analogy to the derivation of the path integral in coordinate representation outlined above. Here we sketch the main steps following another approach that is close to the formulation of Ref. \cite{prokofev_1998}.
We start with a general many-body Hamiltonian with arbitrary pair interaction in second quantization
\begin{equation}
\op{H}=\sum_{i,j}h_{ij}\op{a}^\dagger_i\op{a}_j+\sum_{i<j,k<l}w_{ijkl}^-\creationop_i\creationop_j\annihilop_l\annihilop_k=\op{H}_0+\op{W}\quad\text{with}\quad w^-_{ijkl}:=w_{ijkl}-w_{ijlk},
\end{equation}
where $h_{ij}$ and $w_{ijkl}$ denote the one-particle and two-particle integrals in an arbitrary one-particle basis $\ket{i}$. We split $\op{H}$ uniqely into a diagonal and an off-diagonal part
\begin{equation}
\braket{\occconfig{n_i}|\op{H}|\occconfig{n_j}}=\begin{cases}
\braket{\occconfig{n_i}|\op{D}|\occconfig{n_i}}=D_{\occconfig{n_i}},\qquad&\text{if}\qquad i=j \\
\braket{\occconfig{n_i}|\op{Y}|\occconfig{n_j}}=Y_{\occconfig{n_i},\occconfig{n_j}},\qquad&\text{if}\qquad i\neq j
\end{cases}\;,
\end{equation}
where the matrix elements are given by the Slater-Condon rules \cite{slater-condon}
\begin{align}
\braket{\occconfig{n}|\op{D}|\occconfig{n}}&=\sum_{i} h_{ii}n_i+\sum_{i<j} w_{ijij}^- n_i n_j,\\
\braket{\occconfig{n}|\op{Y}|\occconfig{\bar{n}}}&=\begin{cases}
\displaystyle \Big(h_{pq}+\sum_{i\neq p,q} w_{ipiq}^-n_i\Big)(-1)^{\sum_{m=\min(p,q)+1}^{\max(p,q)-1}n_m} ,&  \occconfig{n}=\occconfig{\bar{n}}_{{q}}^{{p}}\\[0.2cm]
w_{pqrs}^- (-1)^{\sum_{m=p}^{q-1}n_m+\sum_{m=r}^{s-1}\bar{n}_m},&  \occconfig{n}=\occconfig{\bar{n}}_{{r<s}}^{{p<q}} \\[0.1cm]
{0}, & \text{{else}}
\end{cases}.
\label{off_matrix_elments}
\end{align}
that are non-zero only if the states $\ket{\occconfig{n}}$ and $\occconfig{\bar{n}}$ differ by a one-particle or two-particle excitation from $\ket{q}$ to $\ket{p}$ or from $\ket{r}$ and $\ket{s}$ to $\ket{p}$ and $\ket{q}$, respectively. This makes it possible to define an excitation operator by
\begin{align}
\quad\op{q}(s)&:=\begin{cases}\displaystyle
\left(h_{pq} + \sum^\infty_{\substack{j=0\\j\neq p,q}}w^-_{pjqj}\op{n}_j\right)\creationop_p\annihilop_q \quad&\text{if}\quad s=(p,q)\\[0.8cm]
w_{pqrs}^-\creationop_p\creationop_q\annihilop_r\annihilop_s\quad&\text{if}\quad
s=(p,q,r,s)
\end{cases}\;,
\label{eq:def_kink_op}
\end{align}
for all $p\neq q$ and $r\neq s$ and express $\op{Y}$ in terms of all possible one- and two-particle excitations,
$\op{Y}=\sum_{s}\op{q}(s).$
Note that the action of the excitation operator $\op{q}(s)\ket{\occconfig{n}}=q_{\occconfig{\bar{n}},\occconfig{n}}(s)\ket{\occconfig{\bar{n}}}$ is completely determined by $\ket{\occconfig{n}}$ and $s$ with the resulting state $\ket{\occconfig{\bar{n}}}=\ket{\occconfig{{n}}_{{q}}^{{p}}}$ or  $\ket{\occconfig{\bar{n}}}=\ket{\occconfig{{n}}_{{r<s}}^{{p<q}}}$.
Switching to the interaction picture with $\Ham(t) = \op{D} + \op{Y}(t)$ and $\op{Y}(t) = \e{it\op{D}}\op{Y}\e{-it\op{D}}$ one can write the time evolution operator as ($\op{T}$ denotes the time ordering operator)
\begin{align}
\op{U}(t,t_0)= e^{-i \op{D} (t-t_0)} \op{T} e^{-i\int_{t_0}^t\D t' \op{Y}(t')}.
\end{align}
Its action on the exponential function is given by the Dyson series
\begin{align}
\op{T} e^{-i\int_{t_0}^t\D t' \op{Y}(t')}&= \sum_{K=0}^\infty (-i)^K \int_{t_0}^t \D t_1 \dotsi \int_{t_0}^{t_{K-1}}\D t_K \prod_{j=1}^K \op{Y}(t_j).
\end{align}
As the density operator is proportional to the time evolution operator in imaginary time, we arrive at our final expression by carefully evaluating the repeated action of the excitation operators $\op{q}(s,t)$ on the states in the trace
\begin{align}
Z(\beta)&=
\begin{aligned}[t]&\sum_{K=0,\atop K \neq 1}^{\infty} \sum_{\occconfig{n}}
\sum_{s_1}\sum_{s_2}\ldots\sum_{s_{K-1}}\,
\int\limits_{0}^{\beta} d\tau_1 \int\limits_{\tau_1}^{\beta} d\tau_2 \ldots \int\limits_{\tau_{K-1}}^\beta d\tau_K \quad \times 
\nonumber\\
&(-1)^K \exp{\left\{-\sum_{i=0}^{K} D_{\{n^{(i)}\}} (\tau_{i+1}-\tau_i)\right\}} \prod_{i=1}^{K}
q_{\occconfig{n^{(i)}}\occconfig{n^{(i-1)}}}(s_i)  \end{aligned}\\
&= \sum_{K=0,\atop K \neq 1}^{\infty} \sum_{\occconfig{n}} \sum_{s_1\ldots s_{K-1}} \int^{\prime}\mathrm{d}^K{\tau}\, W(K,\occconfig{n}, s_1,\ldots,s_{K-1},\tau_1,\ldots,\tau_K) \;,
\label{eq:closed_path_part_function}
\end{align}
with $\occconfig{n_0}=\occconfig{n_K}=\occconfig{n}$ and, in the last step, we abbreviated the integral over $\tau=-it$ (the primed integral denotes the time ordering) and introduced the weight $W$. The case $K=1$ is forbidden by $\beta$-periodicity. This formula can be interpreted as a sum over all possible paths of occupation number states in the Fock space in imaginary time $\tau$, as shown in Fig. \ref{fig:sample_path}. In this picture sudden changes  in the occupation numbers (``kinks'') are induced by one or two-particle excitations $s_i$ at the times $\tau_i$.  The weight of each path is uniquely determined by the number of kinks $K$, their times and the affected orbitals.
Expectation values that are given by derivatives of $Z$ are readily obtained from Eq.~(\ref{eq:closed_path_part_function}). In particular, the internal energy is given by
\begin{align}
\braket{\Ham}&= \sum_{K=0,\atop K \neq 1}^{\infty} \sum_{\occconfig{n}} \sum_{s_1\ldots s_{K-1}} \int^{\prime}\mathrm{d}^K{\tau}\,\biggl(\frac{1}{\beta} \sum_{i=0}^K D_{\occconfig{n^{(i)}}}(\tau_{i+1}-\tau_i) -\frac{K}{\beta}\biggr) W, \label{eq:KinkKontinuierlicheEnergie}
\end{align}
where, remarkably, the off-diagonal part of the $\op{H}$ enters only indirectly through the number of kinks $K$. 
\par
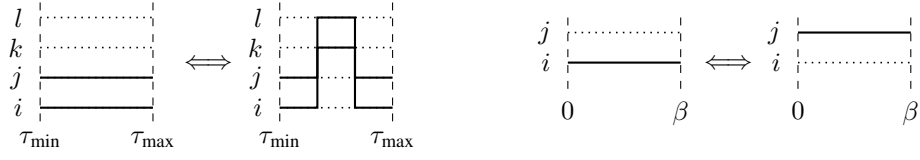
\begin{figure}[t]
\centering
 \input{add_remove_pair_of_kinks.tikz.tex}\qquad\qquad\input{example_1p_excitation.tikz.tex}
 \caption{Left: Add or remove pair of kinks. $\tau_\text{min}$ and $\tau_\text{max}$ correspond to the imaginary times of the neighbouring kinks on the same orbitals. Right: Excite an orbital over the whole $\beta$ range.}
 \label{fig:MC_Steps_1}
\end{figure}
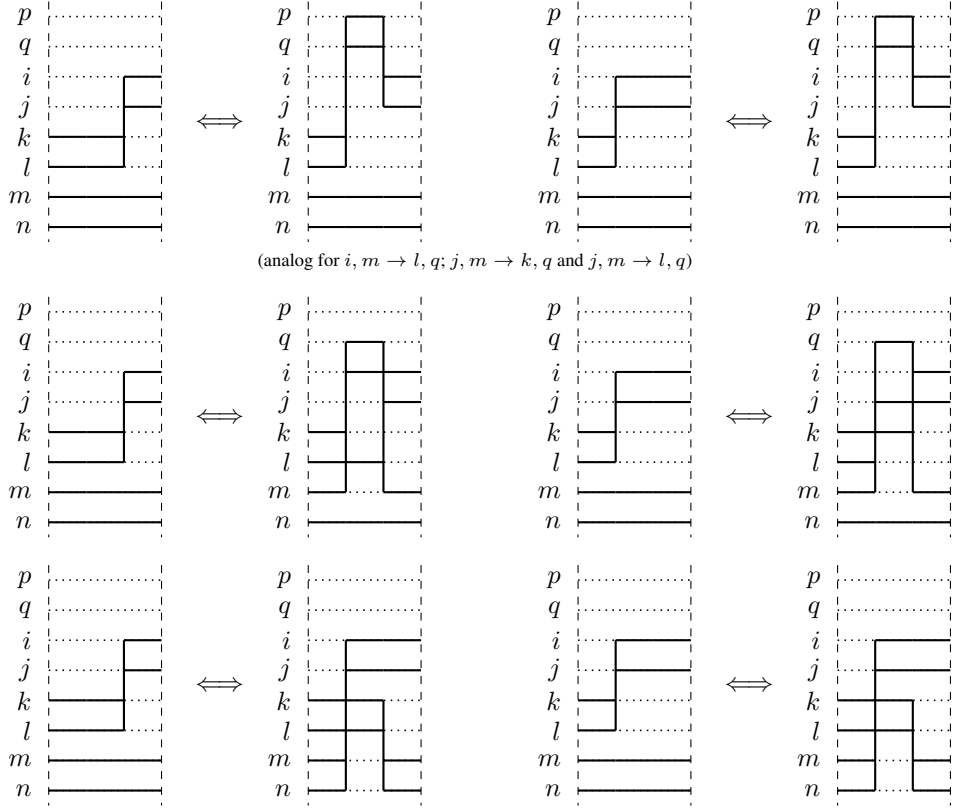
\begin{figure}[t]
\center
 \input{6_possibilities_to_add_and_change_kink.tikz.tex}
 \caption{All possibilities to add or remove a kink. From top to bottom rows correspond to cases i. to iii. The intervals are determined by neighbouring kinks on the affected orbitals.}
 \label{fig:MC_Steps_2}
\end{figure}

So far these expressions are exact. 
For the actual computations a finite number of basis functions $N_B$ has to be chosen. This approximation introduces a basis set incompleteness error, and the convergence to the complete basis set limit has to be carefully investigated. Additionally there is a theoretical limit in the number of kinks that can be stored in memory, but as the FSP limits calculations to a few hundred kinks (see below), this limit is not of any practical relevance. 
To perform these high dimensional integrals and summations we implemented a Metropolis MC scheme. For a general Hamiltonian, a large number of quite complicated Monte Carlo steps is necessary to ensure ergodicity. Details on the general algorithm will be published elsewhere. In the case of the HEG we choose plane waves as underlying one-particle basis. These functions coincide with the eigenfunctions of the interaction-free Hamiltonian, the Hartree-Fock basis functions and the natural orbitals. In this basis the Hamiltonian, $\Ham =\Ham_{\mathrm{el}}+\Ham_{\mathrm{back}}+\Ham_{\mathrm{el-back}}$, can be written as
\begin{align}
\Ham	&= \frac{\hbar^2}{2m} \sum_{\vec{k}}\vec{k}^2 \creationop_{\vec{k}}\annihilop_{\vec{k}} + \frac{1}{2}\frac{4\pi e^2}{V}\sum_{\substack{\vec{k}_i\vec{k}_j\vec{k}_k\vec{k}_l\\\vec{k}_i\neq\vec{k}_k}} \delta_{\vec{k}_i+\vec{k}_j, \vec{k}_k + \vec{k}_l} \frac{1}{(\vec{k}_i-\vec{k}_k)^2} \creationop_{\vec{k}_i}\creationop_{\vec{k}_j} \annihilop_{\vec{k}_l} \annihilop_{\vec{k}_k} + E_M,
\end{align}
where the $\vec{k}_i=\vec{k}_k$ components cancel with the interactions of the positive background and the Madelung energy $E_M$ accounts for the self-interaction of the Ewald summation in periodic boundary conditions. Due to momentum conservation all one-particle excitation operators $\op{q}(i,j)$ vanish and only a subset of MC steps is needed that are sketched below, cf. Figs. \ref{fig:MC_Steps_1} and \ref{fig:MC_Steps_2}.

\begin{enumerate}
  \item Add a pair of kinks:
a) At a random imaginary time $\tau_a$, select two occupied orbitals with the plane wave vectors $\vec{k}_i$ and $\vec{k}_j$. 
b) A random excitation vector $\vec{q}$ is chosen with $\norm{\vec{q}\,}\leq\norm{\vec{q}_\text{max}}$. It is sufficient to set $\norm{\vec{q}_\text{max}}$ to the minimal distance between two $\vec{k}$-vectors, resulting in 6 possible vectors. The step is rejected if one of the new orbitals $\vec{k}_n=\vec{k}_i+\vec{q}$ and $\vec{k}_m=\vec{k}_i-\vec{q}$ is occupied.
c) Using a heat-bath sampling method, the time $\tau_b$ for the second kink is chosen in the interval given by neighbouring kinks or in the whole $\beta$ range if no kinks are present. 
d) If accepted, the kink-pair $(n,m,i,j)$ and $(i,j,n,m)$ will be inserted at $\tau_a$ and $\tau_b$.
%
\item Remove pair of kinks:
a) choose a random kink $s_a$, 
b) choose second kink $s_b$, before or after $s_a$. Reject the step if the kinks do not form a pair of kinks.
c) If accepted, the kinks will be removed.
\item Add one kink
a) A random kink $s_a$ is chosen.
b) Two occupied orbitals with $\vec{k}_i$ and $\vec{k}_j$  are chosen randomly before or after the kink.
c) Depending on the kink and the occupied orbitals one of three different cases apply: \label{enum:choose_orbitals}
  \begin{description}
  \item[i] The kink creates or annihilates particles in both orbitals: randomly choose excitation vector $\vec{q}$. Reject if one of the new orbitals, $\vec{k}_n=\vec{k}_i+\vec{q}$, or $\vec{k}_m=\vec{k}_i-\vec{q}$, is occupied. 
  \item[ii] Only one of the occupied orbitals is affected by the kink: choose an orbital $\vec{k}_n$ from the two unoccupied orbitals that are affected by the kink. The last orbital is determined by $\vec{k}_m=\vec{i}+\vec{j}-\vec{k}_n$. Reject if this orbital is occupied.
  \item[iii] Otherwise both new orbitals $\vec{k}_n$ and  $\vec{k}_m$ are set to the orbitals of the annihilation or creation operators of the kink $s_a$. Reject if the particle excitation does not conserve momentum.
  \end{description}
d) In an interval determined by neighbouring kinks, the time $\tau$ for the new kink is chosen using a heat-bath method. 
e) If accepted add a kink $s_b=(i,j,n,m)$ or $s_b=(n,m,i,j)$ at $\tau$ and change kink $s_a$ accordingly.
%
\item Remove a kink:
a) choose random kink $s_a$.
b) This kink determines a set of kinks that can be removed while changing $s_a$. Choose $s_b$ from these kinks. Reject if the changed kink $s_a'$ does not fulfill momentum conservation or is removed during the process.
c) If accepted, remove $s_b$ and alter $s_a$ accordingly.
\item Change two kinks:
a) Choose a kink $s_a$ randomly.
b) Choose two occupied orbitals $\vec{k}_i$ and $\vec{k}_j$ before or after $s_a$.
c) Determine two unoccupied orbitals $\vec{k}_n$ and  $\vec{k}_m$ analogously to \ref{enum:choose_orbitals}.
d) These orbitals determine a set of kinks that can be changed together with $s_a$. Choose $s_b$ from this set.
e) If the step is accepted, the particles in $\vec{k}_i$ and $\vec{k}_j$ are excited to $\vec{k}_n$ and  $\vec{k}_m$ and the appropriate changes are applied to both kinks.
%
\item Excite whole orbital: 
a) Choose an occupied orbital  $\vec{k}_i$ and an unoccupied orbital $\vec{k}_j$ that are free of any kinks and
b) propose to invert the occupation number of both orbitals.
\end{enumerate} 
\begin{figure}[t]
\centering
 \includegraphics[width=0.49\textwidth]{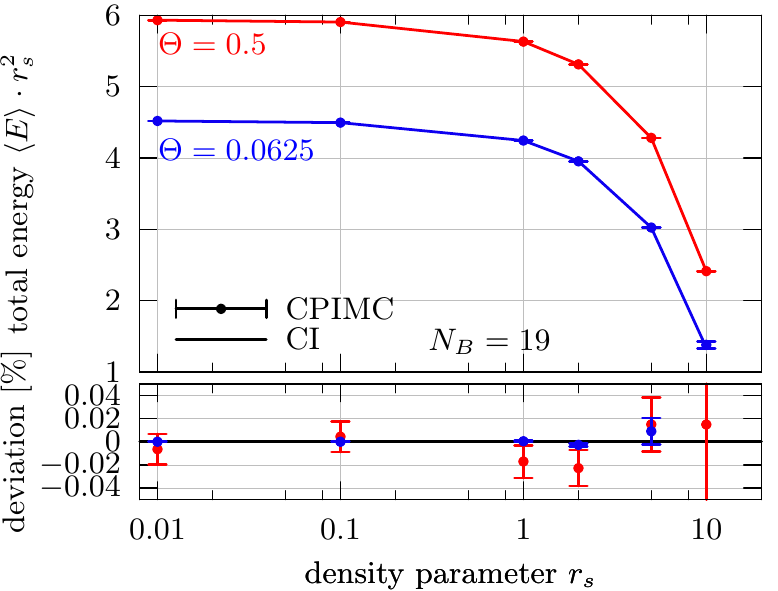}
  \includegraphics[width=0.49\textwidth]{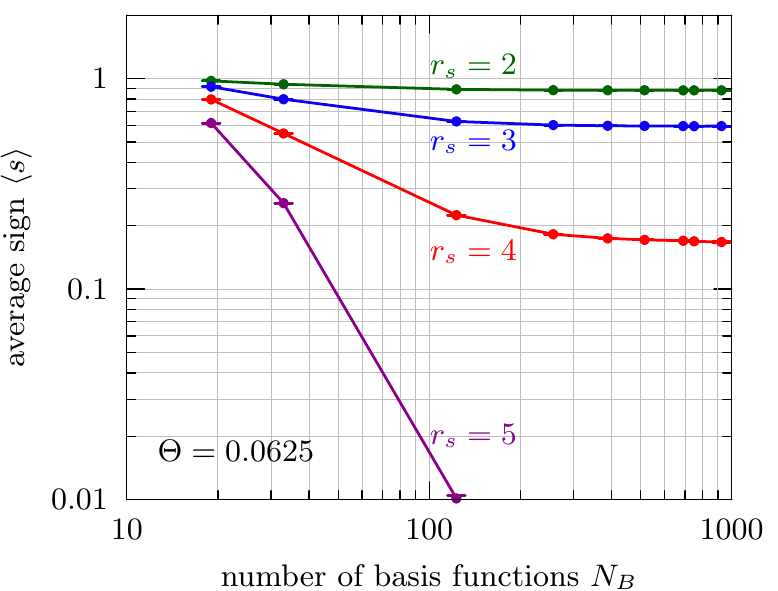}
 \caption{Left: Total energy vs. $r_s$ for two temperatures. CPIMC results (points with error bars) are compared to exact diagonalization results (CI) for the same basis size of $N_B=19$. The lower part shows the relative deviations. Error bars show a one-fold standard deviation. Right: Average sign versus number of basis functions for $\Theta=0.0625$.}
 \label{fig:etot_rs_nb}
\end{figure}
\section{Finite temperature CPIMC results for $N=4$ spin polarized electrons}\label{s:cpimc_ueg}
To demonstrate the validity of the method and its implementation we compare our results to finite temperature configuration interaction (exact diagonalization, CI) results. Because the computational costs grow exponentially with system size, CI calculations are limited to very small numbers of particles and basis functions. It is clear that these results are dominated by finite size effects and are of limited physical value for the uniform electron gas, but their comparison constitutes a rigorous test for CPIMC, as both methods are free of any further approximation and should be numerically identical within statistical errors, if the same basis set is used. This is verified in Fig.~\ref{fig:etot_rs_nb} where the total energy of $N=4$ particles in $N_B=19$ basis functions is shown for different $r_s$ values and temperatures. The error bars correspond to a one-fold standard deviation and demonstrate perfect agreement for all parameters.  For a CPU time of just 1 hour the relative error is as low as $10^{-7}$, for the highest densities and low $T$. At high densities the error is larger for higher $T$ because of the increased thermal fluctuations. At low densities the main source of the statistical error is the FSP, which is more severe for low $T$.
\par
\begin{figure}[t]
\centering
  \includegraphics[width=1\textwidth]{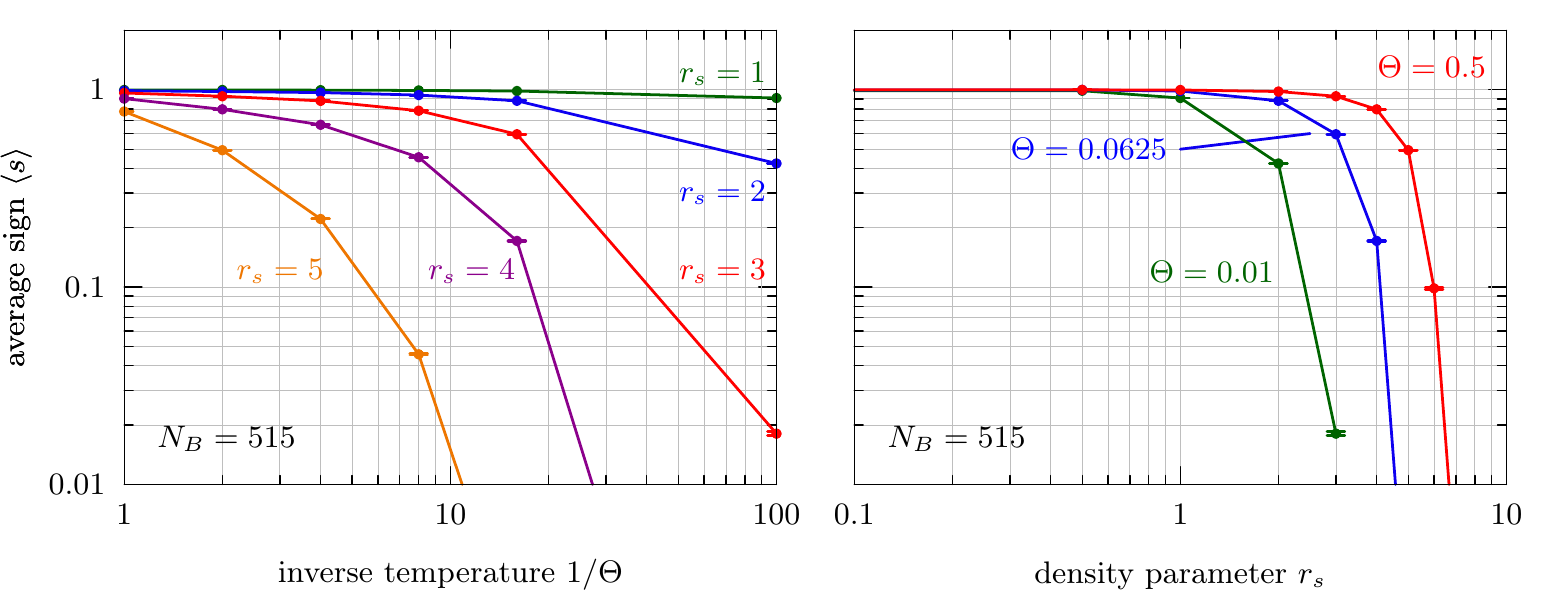}
 \caption{{\bf Left (Right)}: Average sign versus temperature (Brueckner parameter) for $N_B=515$.}
 \label{fig:sign_vs_t_rs}
\end{figure}
\begin{figure}[t]
\begin{minipage}{0.6\textwidth}
 \includegraphics[width=1\textwidth]{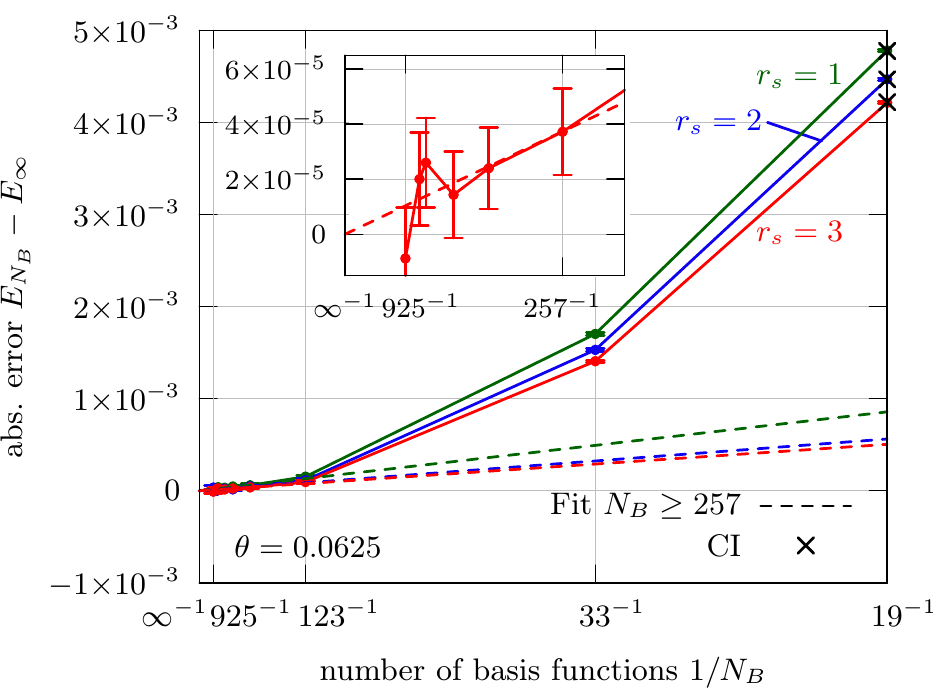}
 \caption{Basis-size incompleteness error of the total energy vs. $N_B$ at temperature $\Theta=0.0625$. Dashed lines are linear extrapolations to $N_B \to \infty$. Error bars correspond to CPIMC runs with a duration of 12 CPU hours. CI results (crosses) are available only for $N_B \le 19$. The inset shows the region used for fitting (for the example $r_s=3$).}
 \label{fig:delta_e_vs_Nb}
\end{minipage}
 \hfil
\begin{minipage}{0.35\textwidth}
\small
 \setfloattype{table}
 \caption{Converged total energy.}
\begin{tabular}{|c|c|c|}
\toprule
   {$\Theta$} &  {$r_s$} &  {$\phantom{-}E/N$ [Ryd]} \\
\midrule
 0.0625 &  0.5 &  15.316652(20) \\
        &    1 &   3.130643(13) \\
        &    2 &   0.429597(10) \\
        &    3 &   0.032051(12) \\
        &    4 &    -0.07229(6) \\
        &    5 &     -0.107(16) \\
   0.25 &  0.5 &     16.2125(7) \\
        &    1 &    3.34891(20) \\
        &    2 &     0.48186(6) \\
        &    3 &     0.05465(4) \\
        &    4 &  -0.059892(34) \\
        &    5 &   -0.09678(23) \\
      1 &  0.5 &    36.3421(30) \\
        &    1 &      8.3856(7) \\
        &    2 &    1.74066(18) \\
        &    3 &     0.61353(9) \\
        &    4 &     0.25383(6) \\
        &    5 &     0.10353(6) \\
\bottomrule
\end{tabular}
\label{tab:1}
\end{minipage}
\end{figure}

To further investigate the FSP we analyze the  dependence of the average sign $\braket{s}$ on the the different parameters. In the left part of Fig.~\ref{fig:sign_vs_t_rs}. $\braket{s}$ is plotted versus $\Theta$. 
As for PIMC the sign decreases exponentially with $1/T$, whereas the dependence on $N_B$ does not have a correspondence in coordinate space. Unfortunately, it can be strong and poses a difficulty for finding the complete basis set limits of the observables. For high densities and moderate $T$, $\braket{s}$ converges and allows for a favorable scaling with $N_B$ which, in the current implementation, is linear, cf. Fig.~\ref{fig:delta_e_vs_Nb}. The dependence of $\braket{s}$ on the density is shown in right part of Fig.~\ref{fig:sign_vs_t_rs}. There is no FSP at all in the high density, interaction-free limit. With decreasing density the sign starts dropping very fast, at a $T$-dependent threshold. The higher the temperature, the lower the density where calculations are feasible. This behavior is complementary to PIMC in coordinate space, which yields accurate results for low densities while suffering from the FSP at high densities.
Due to this complementarity with respect to the FSP there exists a density range where neither PIMC nore CPIMC have a sufficiently large average sign, for larger particle numbers. This makes a direct comparison between CPIMC and (R)PIMC difficult.  In Tab.~\ref{tab:1} we, therefore, present results for $N=4$ particles, which is the lowest particle number for which all MC steps described in Sec.~\ref{s:cpimc} occur, and still has an acceptable average sign for $r_s \le 5$. Our results have been extrapolated to the complete basis set limit by a linear fit as shown in Fig.~\ref{fig:delta_e_vs_Nb} and are considered exact within the given statistical error. The extrapolation assumes a linear convergence over $1/N_B$ for sufficiently large $N_B$, as it was found for the ground state HEG in~\cite{shepherd_2012_a} and is in good agreement also for higher temperatures. We expect that system should also be accessible to direct PIMC in coordinate space, so this appears to be a very useful test system.
%
%

To summarize, this paper presented the first application of CPIMC to the HEG at finite temperatures. Our algorithm yields perfect agreement with CI results for small particle numbers and basis sizes, for a large range of densities and temperatures. For $N=4$ particles at high and moderate degeneracy it has been demonstrated that an accurate extrapolation to the complete basis set limit is possible with small error bars (we underline that this is not possible with CI). Our results can serve as a benchmark for other first-principle methods like (R)PIMC. The FSP of the method has been investigated and found to be qualitatively similar to earlier findings for fermions in a harmonic trap~\cite{schoof_cpp_11}. The complementary dependence of the average sign on the density compared to PIMC in coordinate space allows to reduce the parameter range where the FSP prohibits accurate ab-initio calculations for the HEG. 
More results for larger particle number and different spin polarizations will be presented elsewhere.

\begin{acknowledgement}
 This work was supported by the Deutsche Forschung Gemeinschaft via grant BO1366-10 and the Northern German Supercomputing Alliance (HLRN) via grant shp006.
\end{acknowledgement}


\end{document}

%% file: mydefinitions.tex
\usepackage{printlen}

\DeclarePairedDelimiter\abs{\lvert}{\rvert}
\DeclarePairedDelimiter\norm{\lVert}{\rVert}
\DeclarePairedDelimiter\mittelwert{\langle}{\rangle}
\DeclarePairedDelimiter\klammern{(}{)}
\DeclarePairedDelimiter\kommutator{[}{]}

\DeclarePairedDelimiter\braces{\{}{\}}

\DeclareMathOperator{\Tr}{Tr}

\newcommand{\D}{\ensuremath{\mathrm{d}}}

\newcommand{\op}[1]{\ensuremath{\hat{#1}}}

\newcommand{\Ham}{\op{H}}

\newcommand{\ladderup}{\ensuremath{\op{a}^\dagger}}
\newcommand{\creationop}{\ladderup}
\newcommand{\ladderdown}{\ensuremath{\op{a}^{\vphantom{\dagger}}}}
\newcommand{\annihilop}{\ladderdown}
\newcommand{\occconfig}[1]{\ensuremath{{ \{ #1 \} }}}

\newcommand{\matrixarrows}[1]{{\ensuremath{\mathchoice{%
\mathmakebox[0.2em][l]{\vec{#1}}\reflectbox{$\displaystyle\vec{\phantom{#1}}$}\kern-0.2em}{%
\mathmakebox[0.2em][l]{\vec{#1}}\reflectbox{$\textstyle\vec{\phantom{#1}}$}\kern-0.2em}{%
\mathmakebox[0.15em][l]{\vec{#1}}\reflectbox{$\scriptstyle\vec{\phantom{#1}}$}\kern-0.15em}{%
\mathmakebox[0.1em][l]{\vec{#1}}\reflectbox{$\scriptscriptstyle\vec{\phantom{#1}}$}\kern-0.1em}}}}

\newcommand{\forloop}[5][1]{%
\setcounter{#2}{#3}%
\ifthenelse{#4}{#5\addtocounter{#2}{#1}%
\forloop[#1]{#2}{\value{#2}}{#4}{#5}}%
{}}
\newcounter{crcounter}

\providecommand{\abs}[1]{\ensuremath{\abs{#1}}}
\providecommand{\mittelwert}[1]{\ensuremath{\mittelwert{#1}}}
\providecommand{\klammern}[1]{\ensuremath{\klammern{#1}}}
\providecommand{\kommutator}[1]{\ensuremath{\kommutator{#1}}}

\providecommand{\braces}[1]{\ensuremath{\braces{#1}}}
\providecommand{\text}[1]{\ensuremath{\text{#1}}}
\providecommand{\intertext}[1]{\ensuremath{\intertext{#1}}}

\providecommand{\braces}[1]{\ensuremath{\left\{#1\right\}}}
\providecommand{\braket}[1]{\ensuremath{\braket{#1}}}
\providecommand{\bra}[1]{\ensuremath{\bra{#1}}}
\providecommand{\ket}[1]{\ensuremath{\ket{#1}}}
\providecommand{\binom}[2]{\binom{#1}{#2}}

\providecommand{\SI}[2]{\SI{#1}{#2}}
\providecommand{\num}[1]{\num{#1}}

%% file: original_to_wa_picture_path.tikz.tex
\begin{tikzpicture}[xscale=1.15, yscale=0.5]
\newcommand{\xrange}{10}
\newcommand{\yrange}{1}
\newcommand{\taueins}{\xrange*0.2}
\newcommand{\tauzwei}{\xrange*0.25}
\newcommand{\taudrei}{\xrange*0.4}
\newcommand{\tauvier}{\xrange*0.7}
\newcommand{\taufuenf}{\xrange*0.9}
\newcommand{\zero}{\yrange*1}
\newcommand{\eins}{\yrange*2}
\newcommand{\zwei}{\yrange*3}
\newcommand{\drei}{\yrange*4}
\newcommand{\vier}{\yrange*5}
\newcommand{\fuenf}{\yrange*6}

\draw[->] (0,0) -- +(\xrange+0.5*\xrange/20,0) coordinate (xlabel);
\draw[->] (0,0) -- +(0,\yrange*7) coordinate (ylabel);
\foreach \i in {0,...,5} {
	\draw (-0.1,\i*\yrange+\yrange) node[left] {$\i$} -- (0.1,\i*\yrange+\yrange);
}
\foreach \i/\l in {\taueins/$\tau_1$,\tauzwei/$\tau_2$,\taudrei/$\tau_3$,\tauvier/$\tau_4$,\taufuenf/$\tau_5$} {
\draw (\i,0.1) -- (\i,-0.1) node[below] {\l};
}
\draw (0,0.1) -- (0,-0.1) node[below] {$0$};
\draw (\xrange,0.1) -- (\xrange,-0.1) node[below] {$\beta$};
\node at (0.5*\xrange,-1.7) {imaginary time $\tau$};
\node[rotate=90] at (-0.8,3.5*\yrange) {orbital $i$};

\foreach \i in {1,...,6} {
\draw[semithick,dotted] (0,\i*\yrange) -- (\xrange,\i*\yrange);
}

\draw[dashed] (\taudrei + \tauvier/2.0-\taudrei/2.0 -0.25,0.5) rectangle (\taudrei + \tauvier/2-\taudrei/2+0.25,\yrange*6.5);
\node at (\taudrei + \tauvier/2.0-\taudrei/2.0 ,\fuenf+\yrange) {$\ket{\occconfig{n_3}}=\ket{1010010\ldots}$};

\begin{scope}[thick]
\draw (0,\zero) -| (\tauzwei,\eins) -| (\taudrei,\zero) -- (\xrange,\zero);
\draw (0,\eins) -| (\taueins,\drei) -| (\taudrei,\zwei) -| (\tauvier,\vier) -- (\xrange,\vier);
\draw (0,\vier) -| (\taueins,\fuenf) -| (\taufuenf,\eins) -- (\xrange,\eins);
\draw (\taueins, \drei) -- (\taueins, \vier);
\draw (\taudrei, \eins) -- (\taudrei, \zwei);
 \end{scope}
 
 \node at (\taueins, \fuenf+\yrange) { $s_1 =(3,5,1,4)$};
 \node at (\taufuenf, \fuenf+\yrange) { $s_5 =(1,5)$};

\end{tikzpicture}

%% file: add_remove_pair_of_kinks.tikz.tex
\begin{tikzpicture}[baseline=-0.5ex, xscale=0.5, yscale=0.4]
	\node at (-0.7,0) {};
	\renewcommand{\ll}{1.5}
	\newcommand{\iii}{0.5}
	\newcommand{\kk}{-0.5}
	\newcommand{\jj}{-1.5}
	\draw[dashed] (0,2) -- (0,-2) node[below] {$\tau_\text{min}$};
	\draw[dashed] (3,2) -- (3,-2) node[below] {$\tau_\text{max}$};
	\draw[semithick, dotted] (0,\jj) node[left,xshift=-0.1cm] {$i$} -- (3,\jj);
	\draw[semithick, dotted] (0,\iii) node[left,xshift=-0.1cm] {$k$} -- (3,\iii);
	\draw[semithick, dotted] (0,\kk) node[left,xshift=-0.1cm] {$j$} -- (3,\kk);
	\draw[semithick, dotted] (0,\ll) node[left,xshift=-0.1cm] {$l$} -- (3,\ll);
	\draw[thick] (0,\kk) -- (3,\kk);
	\draw[thick] (0,\jj) -- (3,\jj);
\end{tikzpicture}%
$\Longleftrightarrow$
\begin{tikzpicture}[baseline=-0.5ex, xscale=0.5, yscale=0.4]
	\node at (-0.7,0) {};
	\renewcommand{\ll}{1.5}
	\newcommand{\iii}{0.5}
	\newcommand{\kk}{-0.5}
	\newcommand{\jj}{-1.5}
	\draw[dashed] (0,2) -- (0,-2) node[below] {$\tau_\text{min}$};
	\draw[dashed] (3,2) -- (3,-2) node[below] {$\tau_\text{max}$};
	\draw[semithick, dotted] (0,\jj) node[left,xshift=-0.1cm] {$i$} -- (3,\jj);
	\draw[semithick, dotted] (0,\iii) node[left,xshift=-0.1cm] {$k$} -- (3,\iii);
	\draw[semithick, dotted] (0,\kk) node[left,xshift=-0.1cm] {$j$} -- (3,\kk);
	\draw[semithick, dotted] (0,\ll) node[left,xshift=-0.1cm] {$l$} -- (3,\ll);
	\draw[thick] (0,\kk) -| (1,\ll) -- (2,\ll) |- (3,\kk);
	\draw[thick] (0,\jj) -| (1,\iii) -- (2,\iii) |- (3,\jj);
\end{tikzpicture}%

%% file: example_1p_excitation.tikz.tex
  \begin{tikzpicture}[baseline=-0.5ex,xscale=0.5, yscale=0.4]
     \node at (-0.7,0) {};
     \draw[dashed] (0,1.5) -- (0,-1.0) node[below] {$0$};
     \draw[dashed] (3,1.5) -- (3,-1.0) node[below] {$\beta$};
     \draw[thick] (0,0) node [left,xshift=-0.1cm] {$i$} -- (3,0);
     \draw[semithick,dotted] (0,1) node [left,xshift=-0.1cm] {$j$} -- (3,1);
  \end{tikzpicture}
$\Longleftrightarrow$ 
  \begin{tikzpicture}[baseline=-0.5ex,xscale=0.5, yscale=0.4]
     \node at (-0.7,0) {};
     \draw[dashed] (0,1.5) -- (0,-1.0) node[below] {$0$};
     \draw[dashed] (3,1.5) -- (3,-1.0) node[below] {$\beta$};
     \draw[semithick,dotted] (0,0) node [left,xshift=-0.1cm] {$i$} -- (3,0);
     \draw[thick] (0,1) node [left,xshift=-0.1cm] {$j$} -- (3,1);
  \end{tikzpicture}

%% file: 6_possibilities_to_add_and_change_kink.tikz.tex
\begin{tikzpicture}[baseline=-0.5ex,xscale=0.5, yscale=0.4]
	\node at (-1.1,0) {};
	\newcommand{\pp}{3.5}
	\newcommand{\qq}{2.5}
	\newcommand{\iii}{1.5}
	\newcommand{\jj}{0.5}
	\newcommand{\kk}{-0.5}
	\newcommand{\llll}{-1.5}
	\newcommand{\mmm}{-2.5}
	\newcommand{\nnn}{-3.5}
	\draw[dashed] (0,4) -- (0,-4) ;
	\draw[dashed] (3,4) -- (3,-4) ;
	\draw[semithick, dotted] (0,\jj) node[left,xshift=-0.1cm] {$j$} -- (3,\jj);
	\draw[semithick, dotted] (0,\iii) node[left,xshift=-0.1cm] {$i$} -- (3,\iii);
	\draw[semithick, dotted] (0,\kk) node[left,xshift=-0.1cm] {$k$} -- (3,\kk);
	\draw[semithick, dotted] (0,\llll) node[left,xshift=-0.1cm] {$l$} -- (3,\llll);
	\draw[semithick, dotted] (0,\mmm) node[left,xshift=-0.1cm] {$m$} -- (3,\mmm);
	\draw[semithick, dotted] (0,\nnn) node[left,xshift=-0.1cm] {$n$} -- (3,\nnn);
	\draw[semithick, dotted] (0,\pp) node[left,xshift=-0.1cm] {$p$} -- (3,\pp);
	\draw[semithick, dotted] (0,\qq) node[left,xshift=-0.1cm] {$q$} -- (3,\qq);
	\draw[thick] (0,\llll) -- (1,\llll);
	\draw[thick] (0,\kk) -- (1,\kk);
	\draw[thick] (0,\mmm) -- (1,\mmm);
	\draw[thick] (0,\nnn) -- (1,\nnn);
	\draw[thick] (1,\kk) -- (2,\kk);
	\draw[thick] (1,\llll) -- (2,\llll);
	\draw[thick] (1,\mmm) -- (2,\mmm);
	\draw[thick] (1,\nnn) -- (2,\nnn);
	\draw[thick] (2,\iii) -- (3,\iii);
	\draw[thick] (2,\jj) -- (3,\jj);
	\draw[thick] (2,\nnn) -- (3,\nnn);
	\draw[thick] (2,\mmm) -- (3,\mmm);
	\draw[thick] (2,\llll) -- (2,\iii);
\end{tikzpicture}
\quad$\Longleftrightarrow$\;
\begin{tikzpicture}[baseline=-0.5ex,xscale=0.5, yscale=0.4]
	\node at (-1.1,0) {};
	\newcommand{\pp}{3.5}
	\newcommand{\qq}{2.5}
	\newcommand{\iii}{1.5}
	\newcommand{\jj}{0.5}
	\newcommand{\kk}{-0.5}
	\newcommand{\llll}{-1.5}
	\newcommand{\mmm}{-2.5}
	\newcommand{\nnn}{-3.5}
	\draw[dashed] (0,4) -- (0,-4) ;
	\draw[dashed] (3,4) -- (3,-4) ;
	\draw[semithick, dotted] (0,\jj) node[left,xshift=-0.1cm] {$j$} -- (3,\jj);
	\draw[semithick, dotted] (0,\iii) node[left,xshift=-0.1cm] {$i$} -- (3,\iii);
	\draw[semithick, dotted] (0,\kk) node[left,xshift=-0.1cm] {$k$} -- (3,\kk);
	\draw[semithick, dotted] (0,\llll) node[left,xshift=-0.1cm] {$l$} -- (3,\llll);
	\draw[semithick, dotted] (0,\mmm) node[left,xshift=-0.1cm] {$m$} -- (3,\mmm);
	\draw[semithick, dotted] (0,\nnn) node[left,xshift=-0.1cm] {$n$} -- (3,\nnn);
	\draw[semithick, dotted] (0,\pp) node[left,xshift=-0.1cm] {$p$} -- (3,\pp);
	\draw[semithick, dotted] (0,\qq) node[left,xshift=-0.1cm] {$q$} -- (3,\qq);
	\draw[thick] (0,\llll) -- (1,\llll);
	\draw[thick] (0,\kk) -- (1,\kk);
	\draw[thick] (0,\mmm) -- (1,\mmm);
	\draw[thick] (0,\nnn) -- (1,\nnn);
	\draw[thick] (1,\pp) -- (2,\pp);
	\draw[thick] (1,\qq) -- (2,\qq);
	\draw[thick] (1,\mmm) -- (2,\mmm);
	\draw[thick] (1,\nnn) -- (2,\nnn);
	\draw[thick] (2,\iii) -- (3,\iii);
	\draw[thick] (2,\jj) -- (3,\jj);
	\draw[thick] (2,\nnn) -- (3,\nnn);
	\draw[thick] (2,\mmm) -- (3,\mmm);
	\draw[thick] (1,\llll) -- (1,\pp);
	\draw[thick] (2,\pp) -- (2,\jj);
\end{tikzpicture}\qquad\qquad
\begin{tikzpicture}[baseline=-0.5ex,xscale=0.5, yscale=0.4]
	\node at (-1.1,0) {};
	\newcommand{\pp}{3.5}
	\newcommand{\qq}{2.5}
	\newcommand{\iii}{1.5}
	\newcommand{\jj}{0.5}
	\newcommand{\kk}{-0.5}
	\newcommand{\llll}{-1.5}
	\newcommand{\mmm}{-2.5}
	\newcommand{\nnn}{-3.5}
	\draw[dashed] (0,4) -- (0,-4) ;
	\draw[dashed] (3,4) -- (3,-4) ;
	\draw[semithick, dotted] (0,\jj) node[left,xshift=-0.1cm] {$j$} -- (3,\jj);
	\draw[semithick, dotted] (0,\iii) node[left,xshift=-0.1cm] {$i$} -- (3,\iii);
	\draw[semithick, dotted] (0,\kk) node[left,xshift=-0.1cm] {$k$} -- (3,\kk);
	\draw[semithick, dotted] (0,\llll) node[left,xshift=-0.1cm] {$l$} -- (3,\llll);
	\draw[semithick, dotted] (0,\mmm) node[left,xshift=-0.1cm] {$m$} -- (3,\mmm);
	\draw[semithick, dotted] (0,\nnn) node[left,xshift=-0.1cm] {$n$} -- (3,\nnn);
	\draw[semithick, dotted] (0,\pp) node[left,xshift=-0.1cm] {$p$} -- (3,\pp);
	\draw[semithick, dotted] (0,\qq) node[left,xshift=-0.1cm] {$q$} -- (3,\qq);
	\draw[thick] (0,\llll) -- (1,\llll);
	\draw[thick] (0,\kk) -- (1,\kk);
	\draw[thick] (0,\mmm) -- (1,\mmm);
	\draw[thick] (0,\nnn) -- (1,\nnn);
	\draw[thick] (1,\iii) -- (2,\iii);
	\draw[thick] (1,\jj) -- (2,\jj);
	\draw[thick] (1,\mmm) -- (2,\mmm);
	\draw[thick] (1,\nnn) -- (2,\nnn);
	\draw[thick] (2,\iii) -- (3,\iii);
	\draw[thick] (2,\jj) -- (3,\jj);
	\draw[thick] (2,\nnn) -- (3,\nnn);
	\draw[thick] (2,\mmm) -- (3,\mmm);
	\draw[thick] (1,\llll) -- (1,\iii);
\end{tikzpicture}
\quad$\Longleftrightarrow$\;
\begin{tikzpicture}[baseline=-0.5ex,xscale=0.5, yscale=0.4]
	\node at (-1.1,0) {};
	\newcommand{\pp}{3.5}
	\newcommand{\qq}{2.5}
	\newcommand{\iii}{1.5}
	\newcommand{\jj}{0.5}
	\newcommand{\kk}{-0.5}
	\newcommand{\llll}{-1.5}
	\newcommand{\mmm}{-2.5}
	\newcommand{\nnn}{-3.5}
	\draw[dashed] (0,4) -- (0,-4) ;
	\draw[dashed] (3,4) -- (3,-4) ;
	\draw[semithick, dotted] (0,\jj) node[left,xshift=-0.1cm] {$j$} -- (3,\jj);
	\draw[semithick, dotted] (0,\iii) node[left,xshift=-0.1cm] {$i$} -- (3,\iii);
	\draw[semithick, dotted] (0,\kk) node[left,xshift=-0.1cm] {$k$} -- (3,\kk);
	\draw[semithick, dotted] (0,\llll) node[left,xshift=-0.1cm] {$l$} -- (3,\llll);
	\draw[semithick, dotted] (0,\mmm) node[left,xshift=-0.1cm] {$m$} -- (3,\mmm);
	\draw[semithick, dotted] (0,\nnn) node[left,xshift=-0.1cm] {$n$} -- (3,\nnn);
	\draw[semithick, dotted] (0,\pp) node[left,xshift=-0.1cm] {$p$} -- (3,\pp);
	\draw[semithick, dotted] (0,\qq) node[left,xshift=-0.1cm] {$q$} -- (3,\qq);
	\draw[thick] (0,\llll) -- (1,\llll);
	\draw[thick] (0,\kk) -- (1,\kk);
	\draw[thick] (0,\mmm) -- (1,\mmm);
	\draw[thick] (0,\nnn) -- (1,\nnn);
	\draw[thick] (1,\pp) -- (2,\pp);
	\draw[thick] (1,\qq) -- (2,\qq);
	\draw[thick] (1,\mmm) -- (2,\mmm);
	\draw[thick] (1,\nnn) -- (2,\nnn);
	\draw[thick] (2,\iii) -- (3,\iii);
	\draw[thick] (2,\jj) -- (3,\jj);
	\draw[thick] (2,\nnn) -- (3,\nnn);
	\draw[thick] (2,\mmm) -- (3,\mmm);
	\draw[thick] (1,\llll) -- (1,\pp);
	\draw[thick] (2,\pp) -- (2,\jj);
\end{tikzpicture}\\
\parbox{1\linewidth}
{
 \vspace{0.1cm}\hfil\scriptsize(analog for $i,m\to l,q$; $j,m\to k,q$ and $j,m\to l,q$)\hfil
}\\[0.3cm]
\begin{tikzpicture}[baseline=-0.5ex,xscale=0.5, yscale=0.4]
	\node at (-1.1,0) {};
	\newcommand{\pp}{3.5}
	\newcommand{\qq}{2.5}
	\newcommand{\iii}{1.5}
	\newcommand{\jj}{0.5}
	\newcommand{\kk}{-0.5}
	\newcommand{\llll}{-1.5}
	\newcommand{\mmm}{-2.5}
	\newcommand{\nnn}{-3.5}
	\draw[dashed] (0,4) -- (0,-4) ;
	\draw[dashed] (3,4) -- (3,-4) ;
	\draw[semithick, dotted] (0,\jj) node[left,xshift=-0.1cm] {$j$} -- (3,\jj);
	\draw[semithick, dotted] (0,\iii) node[left,xshift=-0.1cm] {$i$} -- (3,\iii);
	\draw[semithick, dotted] (0,\kk) node[left,xshift=-0.1cm] {$k$} -- (3,\kk);
	\draw[semithick, dotted] (0,\llll) node[left,xshift=-0.1cm] {$l$} -- (3,\llll);
	\draw[semithick, dotted] (0,\mmm) node[left,xshift=-0.1cm] {$m$} -- (3,\mmm);
	\draw[semithick, dotted] (0,\nnn) node[left,xshift=-0.1cm] {$n$} -- (3,\nnn);
	\draw[semithick, dotted] (0,\pp) node[left,xshift=-0.1cm] {$p$} -- (3,\pp);
	\draw[semithick, dotted] (0,\qq) node[left,xshift=-0.1cm] {$q$} -- (3,\qq);
	\draw[thick] (0,\llll) -- (1,\llll);
	\draw[thick] (0,\kk) -- (1,\kk);
	\draw[thick] (0,\mmm) -- (1,\mmm);
	\draw[thick] (0,\nnn) -- (1,\nnn);
	\draw[thick] (1,\kk) -- (2,\kk);
	\draw[thick] (1,\llll) -- (2,\llll);
	\draw[thick] (1,\mmm) -- (2,\mmm);
	\draw[thick] (1,\nnn) -- (2,\nnn);
	\draw[thick] (2,\iii) -- (3,\iii);
	\draw[thick] (2,\jj) -- (3,\jj);
	\draw[thick] (2,\nnn) -- (3,\nnn);
	\draw[thick] (2,\mmm) -- (3,\mmm);
	\draw[thick] (2,\llll) -- (2,\iii);
\end{tikzpicture}
\quad$\Longleftrightarrow$\;
\begin{tikzpicture}[baseline=-0.5ex,xscale=0.5, yscale=0.4]
	\node at (-1.1,0) {};
	\newcommand{\pp}{3.5}
	\newcommand{\qq}{2.5}
	\newcommand{\iii}{1.5}
	\newcommand{\jj}{0.5}
	\newcommand{\kk}{-0.5}
	\newcommand{\llll}{-1.5}
	\newcommand{\mmm}{-2.5}
	\newcommand{\nnn}{-3.5}
	\draw[dashed] (0,4) -- (0,-4) ;
	\draw[dashed] (3,4) -- (3,-4) ;
	\draw[semithick, dotted] (0,\jj) node[left,xshift=-0.1cm] {$j$} -- (3,\jj);
	\draw[semithick, dotted] (0,\iii) node[left,xshift=-0.1cm] {$i$} -- (3,\iii);
	\draw[semithick, dotted] (0,\kk) node[left,xshift=-0.1cm] {$k$} -- (3,\kk);
	\draw[semithick, dotted] (0,\llll) node[left,xshift=-0.1cm] {$l$} -- (3,\llll);
	\draw[semithick, dotted] (0,\mmm) node[left,xshift=-0.1cm] {$m$} -- (3,\mmm);
	\draw[semithick, dotted] (0,\nnn) node[left,xshift=-0.1cm] {$n$} -- (3,\nnn);
	\draw[semithick, dotted] (0,\pp) node[left,xshift=-0.1cm] {$p$} -- (3,\pp);
	\draw[semithick, dotted] (0,\qq) node[left,xshift=-0.1cm] {$q$} -- (3,\qq);
	\draw[thick] (0,\llll) -- (1,\llll);
	\draw[thick] (0,\kk) -- (1,\kk);
	\draw[thick] (0,\mmm) -- (1,\mmm);
	\draw[thick] (0,\nnn) -- (1,\nnn);
	\draw[thick] (1,\iii) -- (2,\iii);
	\draw[thick] (1,\llll) -- (2,\llll);
	\draw[thick] (1,\qq) -- (2,\qq);
	\draw[thick] (1,\nnn) -- (2,\nnn);
	\draw[thick] (2,\iii) -- (3,\iii);
	\draw[thick] (2,\jj) -- (3,\jj);
	\draw[thick] (2,\nnn) -- (3,\nnn);
	\draw[thick] (2,\mmm) -- (3,\mmm);
	\draw[thick] (1,\mmm) -- (1,\qq);
	\draw[thick] (2,\mmm) -- (2,\qq);
\end{tikzpicture}\qquad\qquad
\begin{tikzpicture}[baseline=-0.5ex,xscale=0.5, yscale=0.4]
	\node at (-1.1,0) {};
	\newcommand{\pp}{3.5}
	\newcommand{\qq}{2.5}
	\newcommand{\iii}{1.5}
	\newcommand{\jj}{0.5}
	\newcommand{\kk}{-0.5}
	\newcommand{\llll}{-1.5}
	\newcommand{\mmm}{-2.5}
	\newcommand{\nnn}{-3.5}
	\draw[dashed] (0,4) -- (0,-4) ;
	\draw[dashed] (3,4) -- (3,-4) ;
	\draw[semithick, dotted] (0,\jj) node[left,xshift=-0.1cm] {$j$} -- (3,\jj);
	\draw[semithick, dotted] (0,\iii) node[left,xshift=-0.1cm] {$i$} -- (3,\iii);
	\draw[semithick, dotted] (0,\kk) node[left,xshift=-0.1cm] {$k$} -- (3,\kk);
	\draw[semithick, dotted] (0,\llll) node[left,xshift=-0.1cm] {$l$} -- (3,\llll);
	\draw[semithick, dotted] (0,\mmm) node[left,xshift=-0.1cm] {$m$} -- (3,\mmm);
	\draw[semithick, dotted] (0,\nnn) node[left,xshift=-0.1cm] {$n$} -- (3,\nnn);
	\draw[semithick, dotted] (0,\pp) node[left,xshift=-0.1cm] {$p$} -- (3,\pp);
	\draw[semithick, dotted] (0,\qq) node[left,xshift=-0.1cm] {$q$} -- (3,\qq);
	\draw[thick] (0,\llll) -- (1,\llll);
	\draw[thick] (0,\kk) -- (1,\kk);
	\draw[thick] (0,\mmm) -- (1,\mmm);
	\draw[thick] (0,\nnn) -- (1,\nnn);
	\draw[thick] (1,\iii) -- (2,\iii);
	\draw[thick] (1,\jj) -- (2,\jj);
	\draw[thick] (1,\mmm) -- (2,\mmm);
	\draw[thick] (1,\nnn) -- (2,\nnn);
	\draw[thick] (2,\iii) -- (3,\iii);
	\draw[thick] (2,\jj) -- (3,\jj);
	\draw[thick] (2,\nnn) -- (3,\nnn);
	\draw[thick] (2,\mmm) -- (3,\mmm);
	\draw[thick] (1,\llll) -- (1,\iii);
\end{tikzpicture}
\quad$\Longleftrightarrow$\;
\begin{tikzpicture}[baseline=-0.5ex,xscale=0.5, yscale=0.4]
	\node at (-1.1,0) {};
	\newcommand{\pp}{3.5}
	\newcommand{\qq}{2.5}
	\newcommand{\iii}{1.5}
	\newcommand{\jj}{0.5}
	\newcommand{\kk}{-0.5}
	\newcommand{\llll}{-1.5}
	\newcommand{\mmm}{-2.5}
	\newcommand{\nnn}{-3.5}
	\draw[dashed] (0,4) -- (0,-4) ;
	\draw[dashed] (3,4) -- (3,-4) ;
	\draw[semithick, dotted] (0,\jj) node[left,xshift=-0.1cm] {$j$} -- (3,\jj);
	\draw[semithick, dotted] (0,\iii) node[left,xshift=-0.1cm] {$i$} -- (3,\iii);
	\draw[semithick, dotted] (0,\kk) node[left,xshift=-0.1cm] {$k$} -- (3,\kk);
	\draw[semithick, dotted] (0,\llll) node[left,xshift=-0.1cm] {$l$} -- (3,\llll);
	\draw[semithick, dotted] (0,\mmm) node[left,xshift=-0.1cm] {$m$} -- (3,\mmm);
	\draw[semithick, dotted] (0,\nnn) node[left,xshift=-0.1cm] {$n$} -- (3,\nnn);
	\draw[semithick, dotted] (0,\pp) node[left,xshift=-0.1cm] {$p$} -- (3,\pp);
	\draw[semithick, dotted] (0,\qq) node[left,xshift=-0.1cm] {$q$} -- (3,\qq);
	\draw[thick] (0,\llll) -- (1,\llll);
	\draw[thick] (0,\kk) -- (1,\kk);
	\draw[thick] (0,\mmm) -- (1,\mmm);
	\draw[thick] (0,\nnn) -- (1,\nnn);
	\draw[thick] (1,\kk) -- (2,\kk);
	\draw[thick] (1,\jj) -- (2,\jj);
	\draw[thick] (1,\qq) -- (2,\qq);
	\draw[thick] (1,\nnn) -- (2,\nnn);
	\draw[thick] (2,\iii) -- (3,\iii);
	\draw[thick] (2,\jj) -- (3,\jj);
	\draw[thick] (2,\nnn) -- (3,\nnn);
	\draw[thick] (2,\mmm) -- (3,\mmm);
	\draw[thick] (1,\mmm) -- (1,\qq);
	\draw[thick] (2,\mmm) -- (2,\qq);
\end{tikzpicture}\\[0.3cm]
\begin{tikzpicture}[baseline=-0.5ex,xscale=0.5, yscale=0.4]
	\node at (-1.1,0) {};
	\newcommand{\pp}{3.5}
	\newcommand{\qq}{2.5}
	\newcommand{\iii}{1.5}
	\newcommand{\jj}{0.5}
	\newcommand{\kk}{-0.5}
	\newcommand{\llll}{-1.5}
	\newcommand{\mmm}{-2.5}
	\newcommand{\nnn}{-3.5}
	\draw[dashed] (0,4) -- (0,-4) ;
	\draw[dashed] (3,4) -- (3,-4) ;
	\draw[semithick, dotted] (0,\jj) node[left,xshift=-0.1cm] {$j$} -- (3,\jj);
	\draw[semithick, dotted] (0,\iii) node[left,xshift=-0.1cm] {$i$} -- (3,\iii);
	\draw[semithick, dotted] (0,\kk) node[left,xshift=-0.1cm] {$k$} -- (3,\kk);
	\draw[semithick, dotted] (0,\llll) node[left,xshift=-0.1cm] {$l$} -- (3,\llll);
	\draw[semithick, dotted] (0,\mmm) node[left,xshift=-0.1cm] {$m$} -- (3,\mmm);
	\draw[semithick, dotted] (0,\nnn) node[left,xshift=-0.1cm] {$n$} -- (3,\nnn);
	\draw[semithick, dotted] (0,\pp) node[left,xshift=-0.1cm] {$p$} -- (3,\pp);
	\draw[semithick, dotted] (0,\qq) node[left,xshift=-0.1cm] {$q$} -- (3,\qq);
	\draw[thick] (0,\llll) -- (1,\llll);
	\draw[thick] (0,\kk) -- (1,\kk);
	\draw[thick] (0,\mmm) -- (1,\mmm);
	\draw[thick] (0,\nnn) -- (1,\nnn);
	\draw[thick] (1,\kk) -- (2,\kk);
	\draw[thick] (1,\llll) -- (2,\llll);
	\draw[thick] (1,\mmm) -- (2,\mmm);
	\draw[thick] (1,\nnn) -- (2,\nnn);
	\draw[thick] (2,\iii) -- (3,\iii);
	\draw[thick] (2,\jj) -- (3,\jj);
	\draw[thick] (2,\nnn) -- (3,\nnn);
	\draw[thick] (2,\mmm) -- (3,\mmm);
	\draw[thick] (2,\llll) -- (2,\iii);
\end{tikzpicture}
\quad$\Longleftrightarrow$\;
\begin{tikzpicture}[baseline=-0.5ex,xscale=0.5, yscale=0.4]
	\node at (-1.1,0) {};
	\newcommand{\pp}{3.5}
	\newcommand{\qq}{2.5}
	\newcommand{\iii}{1.5}
	\newcommand{\jj}{0.5}
	\newcommand{\kk}{-0.5}
	\newcommand{\llll}{-1.5}
	\newcommand{\mmm}{-2.5}
	\newcommand{\nnn}{-3.5}
	\draw[dashed] (0,4) -- (0,-4) ;
	\draw[dashed] (3,4) -- (3,-4) ;
	\draw[semithick, dotted] (0,\jj) node[left,xshift=-0.1cm] {$j$} -- (3,\jj);
	\draw[semithick, dotted] (0,\iii) node[left,xshift=-0.1cm] {$i$} -- (3,\iii);
	\draw[semithick, dotted] (0,\kk) node[left,xshift=-0.1cm] {$k$} -- (3,\kk);
	\draw[semithick, dotted] (0,\llll) node[left,xshift=-0.1cm] {$l$} -- (3,\llll);
	\draw[semithick, dotted] (0,\mmm) node[left,xshift=-0.1cm] {$m$} -- (3,\mmm);
	\draw[semithick, dotted] (0,\nnn) node[left,xshift=-0.1cm] {$n$} -- (3,\nnn);
	\draw[semithick, dotted] (0,\pp) node[left,xshift=-0.1cm] {$p$} -- (3,\pp);
	\draw[semithick, dotted] (0,\qq) node[left,xshift=-0.1cm] {$q$} -- (3,\qq);
	\draw[thick] (0,\llll) -- (1,\llll);
	\draw[thick] (0,\kk) -- (1,\kk);
	\draw[thick] (0,\mmm) -- (1,\mmm);
	\draw[thick] (0,\nnn) -- (1,\nnn);
	\draw[thick] (1,\kk) -- (2,\kk);
	\draw[thick] (1,\llll) -- (2,\llll);
	\draw[thick] (1,\iii) -- (2,\iii);
	\draw[thick] (1,\jj) -- (2,\jj);
	\draw[thick] (2,\iii) -- (3,\iii);
	\draw[thick] (2,\jj) -- (3,\jj);
	\draw[thick] (2,\nnn) -- (3,\nnn);
	\draw[thick] (2,\mmm) -- (3,\mmm);
	\draw[thick] (1,\nnn) -- (1,\iii);
	\draw[thick] (2,\kk) -- (2,\nnn);
\end{tikzpicture}\qquad\qquad
\begin{tikzpicture}[baseline=-0.5ex,xscale=0.5, yscale=0.4]
	\node at (-1.1,0) {};
	\newcommand{\pp}{3.5}
	\newcommand{\qq}{2.5}
	\newcommand{\iii}{1.5}
	\newcommand{\jj}{0.5}
	\newcommand{\kk}{-0.5}
	\newcommand{\llll}{-1.5}
	\newcommand{\mmm}{-2.5}
	\newcommand{\nnn}{-3.5}
	\draw[dashed] (0,4) -- (0,-4) ;
	\draw[dashed] (3,4) -- (3,-4) ;
	\draw[semithick, dotted] (0,\jj) node[left,xshift=-0.1cm] {$j$} -- (3,\jj);
	\draw[semithick, dotted] (0,\iii) node[left,xshift=-0.1cm] {$i$} -- (3,\iii);
	\draw[semithick, dotted] (0,\kk) node[left,xshift=-0.1cm] {$k$} -- (3,\kk);
	\draw[semithick, dotted] (0,\llll) node[left,xshift=-0.1cm] {$l$} -- (3,\llll);
	\draw[semithick, dotted] (0,\mmm) node[left,xshift=-0.1cm] {$m$} -- (3,\mmm);
	\draw[semithick, dotted] (0,\nnn) node[left,xshift=-0.1cm] {$n$} -- (3,\nnn);
	\draw[semithick, dotted] (0,\pp) node[left,xshift=-0.1cm] {$p$} -- (3,\pp);
	\draw[semithick, dotted] (0,\qq) node[left,xshift=-0.1cm] {$q$} -- (3,\qq);
	\draw[thick] (0,\llll) -- (1,\llll);
	\draw[thick] (0,\kk) -- (1,\kk);
	\draw[thick] (0,\mmm) -- (1,\mmm);
	\draw[thick] (0,\nnn) -- (1,\nnn);
	\draw[thick] (1,\iii) -- (2,\iii);
	\draw[thick] (1,\jj) -- (2,\jj);
	\draw[thick] (1,\mmm) -- (2,\mmm);
	\draw[thick] (1,\nnn) -- (2,\nnn);
	\draw[thick] (2,\iii) -- (3,\iii);
	\draw[thick] (2,\jj) -- (3,\jj);
	\draw[thick] (2,\nnn) -- (3,\nnn);
	\draw[thick] (2,\mmm) -- (3,\mmm);
	\draw[thick] (1,\llll) -- (1,\iii);
\end{tikzpicture}
\quad$\Longleftrightarrow$\;
\begin{tikzpicture}[baseline=-0.5ex,xscale=0.5, yscale=0.4]
	\node at (-1.1,0) {};
	\newcommand{\pp}{3.5}
	\newcommand{\qq}{2.5}
	\newcommand{\iii}{1.5}
	\newcommand{\jj}{0.5}
	\newcommand{\kk}{-0.5}
	\newcommand{\llll}{-1.5}
	\newcommand{\mmm}{-2.5}
	\newcommand{\nnn}{-3.5}
	\draw[dashed] (0,4) -- (0,-4) ;
	\draw[dashed] (3,4) -- (3,-4) ;
	\draw[semithick, dotted] (0,\jj) node[left,xshift=-0.1cm] {$j$} -- (3,\jj);
	\draw[semithick, dotted] (0,\iii) node[left,xshift=-0.1cm] {$i$} -- (3,\iii);
	\draw[semithick, dotted] (0,\kk) node[left,xshift=-0.1cm] {$k$} -- (3,\kk);
	\draw[semithick, dotted] (0,\llll) node[left,xshift=-0.1cm] {$l$} -- (3,\llll);
	\draw[semithick, dotted] (0,\mmm) node[left,xshift=-0.1cm] {$m$} -- (3,\mmm);
	\draw[semithick, dotted] (0,\nnn) node[left,xshift=-0.1cm] {$n$} -- (3,\nnn);
	\draw[semithick, dotted] (0,\pp) node[left,xshift=-0.1cm] {$p$} -- (3,\pp);
	\draw[semithick, dotted] (0,\qq) node[left,xshift=-0.1cm] {$q$} -- (3,\qq);
	\draw[thick] (0,\llll) -- (1,\llll);
	\draw[thick] (0,\kk) -- (1,\kk);
	\draw[thick] (0,\mmm) -- (1,\mmm);
	\draw[thick] (0,\nnn) -- (1,\nnn);
	\draw[thick] (1,\iii) -- (2,\iii);
	\draw[thick] (1,\jj) -- (2,\jj);
	\draw[thick] (1,\kk) -- (2,\kk);
	\draw[thick] (1,\llll) -- (2,\llll);
	\draw[thick] (2,\iii) -- (3,\iii);
	\draw[thick] (2,\jj) -- (3,\jj);
	\draw[thick] (2,\nnn) -- (3,\nnn);
	\draw[thick] (2,\mmm) -- (3,\mmm);
	\draw[thick] (1,\nnn) -- (1,\iii);
	\draw[thick] (2,\kk) -- (2,\nnn);
\end{tikzpicture}%

%% file: egas_cpimcxxx.bbl
\begin{thebibliography}{10}

\bibitem{lindl_04} J. D. Lindl {\em et al.}
Phys. Plasmas {\bf 11}, 339 (2004).

\bibitem{hurricane_nif14} O. Hurricane {\em et al.}, Nature {\bf 506}, 346 (2014)

\bibitem{hanson_13} T.J. Awe {\em et al.}, Phys. Rev. Lett. {\bf 111}, 235005 (2013)

\bibitem{hochstuhl_jcp11} D.~Hochstuhl, and M.~Bonitz, J. Chem. Phys. {\bf 134}, 084106 (2011)



\bibitem{mahan-book} G.D. Mahan, {\it Many-Particle Physics}, Plenum 2000

\bibitem{bonitz_cpp13} M. Bonitz, S. Hermanns, and K. Balzer, Contrib. Plasma Phys. {\bf 53}, 778 (2013), arXiv:1309.4574

\bibitem{ceperley95rmp} D.M.~Ceperley,
Rev. Mod. Phys. {\bf 65}, 279 (1995)

\bibitem{filinov-etal.00pla} V.S.~Filinov, V.E.~Fortov, M.~Bonitz, and D. Kremp,
Physics Lett. A {\bf 274}, 228 (2000)

\bibitem{FiBoEbFo01}
V.S.~Filinov, M.~Bonitz, W.~Ebeling, and V.E.~Fortov, Plasma Phys.
Control. Fusion {\bf 43}, 743 (2001)

\bibitem{mil-pol} B.~Militzer, and R.~Pollock, Phys. Rev. E {\bf 61}, 3470 (2000)

\bibitem{brown_prl13} E.W. Brown, B. K. Clark, J. L. DuBois, and D. M.
Ceperley, Phys. Rev. Lett. {\bf 110}, 146405 (2013).

\bibitem{filinov_14} V.S. Filinov, M. Bonitz, Zh. Moldabekov, and V.E. Fortov, submitted for publication, 
arxiv: 1407.3600

\bibitem{brown_prb13} E.W. Brown, J. L. DuBois, M. Holzmann, and D. M.
Ceperley, Phys. Rev. B {\bf 88}, 081102(R) (2013); {\bf 88}, 199901(E) (2013).

\bibitem{karasiev_prl14} V.V. Karasiev, T. Sjostrom, J. Dufty, and S. B. Trickey,
Phys. Rev. Lett. {\bf 112}, 076403 (2014) and Supplementary Material.


\bibitem{schoof_cpp_11} T. Schoof, M. Bonitz, A. Filinov, D. Hochstuhl, and J.W. Dufty, Contrib. Plasma Phys. {\bf 51}, 687 (2011)

\bibitem{simon_springer14} S. Groth, T. Schoof, and M. Bonitz, Chapter in: {\em Complex Plasmas: Scientific Challenges and Technological Opportunities}, M. Bonitz, K. Becker, J. Lopez, and H. Thomsen (eds.), Springer 2014

\bibitem{prokofev_1998} N. V. Prokof’ev, B. V. Svistunov, and I. S. Tupitsyn, J. Exp. Theor. Phys. {\bf 87}, 310 (1998).



\bibitem{slater-condon}  T. Helgaker, P. Jorgensen, and J. Olsen, \emph{Molecular Electronic-Structure
Theory}, (Wiley, Chichester, Hoboken, 2000)

\bibitem{shepherd_2012_a} J. J. Shepherd, G. Booth, A. Grüneis, and A. Alavi, Phys. Rev. B {\bf 85}, 081103 (2012).

\bibitem{shepherd_2012_b} J. J. Shepherd, A. Grüneis, G. H. Booth, G. Kresse, and A. Alavi, Phys. Rev. B {\bf 86}, 035111 (2012).

\end{thebibliography}
